\documentclass[pra, preprint, superscriptaddress]{revtex4}
\usepackage{graphicx}
\usepackage{dcolumn}
\usepackage{amsmath}
\usepackage{amssymb}
\usepackage{color}
\usepackage{float}
\usepackage{epstopdf}

\def\beq{\begin{equation}}
\def\eeq{\end{equation}}
\def\eq#1{(\ref{#1})}
\def\simg{\,\hbox{\kern.1em \lower.6ex \hbox{$\sim$} \kern-1.12em
          \raise.6ex \hbox{$>$} }}
\def\siml{\,\hbox{\kern.1em \lower.6ex \hbox{$\sim$} \kern-1.12em
          \raise.6ex \hbox{$<$} }}
\newcommand{\Table}[4]{
\begin{table}[H]\begin{center}{#3}
\parbox{#2cm}{
\caption[table]{\renewcommand{\baselinestretch}{0.8} \small
                                           \hspace{-0.3truecm}#4}
\label{#1}}
\end{center}
\end{table}
}
\begin{document}
\title{A semiclassical analysis of the Efimov energy spectrum in the unitary limit}
\author{Rajat K. Bhaduri}
\affiliation{Department of Physics and Astronomy, McMaster University, 
Hamilton, Canada L9H 6T6}
\author{Matthias Brack}
\affiliation{Institute for Theoretical Physics, University of Regensburg,
D-93040 Regensburg, Germany}
\author{M. V. N. Murthy}
\affiliation{Institute of Mathematical Sciences, Chennai, 600 113 India}
\date{\today}

\begin{abstract}
We demonstrate that the ($s$-wave) geometric spectrum of the Efimov energy 
levels in the unitary limit is generated by the radial motion of a primitive 
periodic orbit (and its harmonics) of the corresponding classical system.
The action of the primitive orbit depends logarithmically on the energy. 
It is shown to be consistent with an inverse-squared radial potential 
with a lower cut-off radius. The lowest-order WKB 
quantization, including the Langer correction, is shown to reproduce the 
geometric scaling of the energy spectrum. The (WKB) mean-squared radii of 
the Efimov states scale geometrically like the inverse of their energies. 
The WKB wavefunctions, regularized near the classical turning point by
Langer's generalized connection formula, are practically indistinguishable 
from the exact wave functions even for the lowest ($n=0$) state, apart 
from a tiny shift of its zeros that remains constant for large $n$.

\bigskip

     
\end{abstract}

\maketitle

\section{Introduction}

Two particles that are just shy of binding may develop an infinite 
number of shallow bound states when a third particle is added. This was 
predicted by Efimov \cite{efimov} forty years back, and has only been 
recently verified experimentally with an ultra-cold gas of optically 
trapped $^{133}$Cs atoms \cite{kraemer}. Subsequently, Barontini {\it et 
al.} \cite{baront} have have found evidence for two kinds of Efimov 
trimers with $^{41}$K and $^{87}$Rb atoms.  Efimov considered three identical 
bosons interacting pairwise with an interaction whose range $r_0$ is 
much smaller than the interatomic scattering length $a$. Using the 
hyperspherical coordinates for the three-body problem, he showed that 
the effective potential in the hyperradial coordinate $R$ between the length 
scales of $r_0$ and $|a|$ is of an inverse-square type. In the symmetric 
$L=0$ three-body state, this effective interaction is sufficiently 
attractive to give rise to the Efimov spectrum for the trimers.
A signature of the Efimov spectrum is its geometric scaling, with the 
ratio of the adjacent energy levels being constant. This was predicted by 
Efimov \cite{efimov} and has been verified by recent experiments \cite{zacca}. 

Although in experiments the Efimov spectrum can only be measured for large 
but finite scattering lengths $a$, our semiclassical analysis of the geometric 
spectrum in Sec.\ II is done in the unitary limit ($|a|\to\infty$). The geometric 
scaling of the spectrum then holds right up to the accumulation point at $E=0$,  
and the effective potential is of inverse-square type for all distances $R>r_0$. 
This limit itself has interesting properties and given rise to various 
theoretical research \cite{rajeev,bra,grif}. Our contribution to this research 
here is a semiclassical description of the unitary Efimov system.

In the periodic orbit theory (POT) \cite{gutz}, there is an intimate connection 
between classical periodic orbits and the quantum energy spectrum of a system 
through so-called trace formulae (cf.\ \cite{book} for an introduction and 
applications of the POT). In Sec.\ II, we derive an exact trace formula from the 
Efimov spectrum and show that the action of the classical periodic orbit 
generating the quantum spectrum depends logarithmically on the energy. The 
corresponding average level density leads, via an inverse Abel transform, to 
an inverse-squared radial one-body potential that creates the Efimov spectrum 
in the limit $E\to 0$. 

In Sec.\ \ref{secinv}, we first re-derive the quantum-mechanical solutions 
of an inverse-squared potential with a lower cut-off radius $R_c$. The exact 
quantum wave function is a modified Bessel function of imaginary order 
\cite{rajeev}. Next we calculate the first-order WKB wave function (including 
the Langer correction \cite{lang}). We show that its leading term in the 
classically allowed region is identical with the leading term of the 
exact wave function for high-lying states ($n\gg1$), responsible for the 
geometric scaling of the spectrum. In the classically forbidden region, 
it decays exponentially like the exact one. The WKB eigenvalues are obtained 
by quantizing the classical action of the inverse-squared potential, whose 
leading term is the action appearing in the semiclassical trace formula 
discussed in Sec.\ \ref{secgeo}. We find that the WKB spectrum, although 
shlightly phase shifted with respect to the exact one, reproduces the 
geometric scaling of the quantum-mechanical energies and mean-squared radii 
with the same scaling factor. Remarkably, the WKB wave functions, when regularized 
near the classical turning point by Langer's generalized connection formula 
\cite{lang}, are -- apart from their slightly shifted zeros -- practically 
indistinguishable from the quantum-mechanical wave functions even for the 
lowest state ($n=0$).
  

\section{Geometric spectrum}
\label{secgeo}

In the unitary limit, three identical bosons in the symmetric $s$ state ($L=0)$, 
have the energy spectrum given by 
\beq 
E_n=E_0 \exp(-2\pi n/s_0)\,, \quad\qquad n=0,1,2,\dots,\infty
\label{spectrum}
\eeq
so that $E_{n+1}/E_n$ is constant and independent of $n$ and $E_0$. This is called 
the geometric spectrum. $E_0<0$ and $s_0>0$ are constants that depend on the system. 
For three equal-mass bosons, one has   
\beq
\exp (\pi/s_0)=22.694 \qquad \Rightarrow \qquad s_0=1.00624\,.
\label{space}
\eeq
The energy $E_0$, which corresponds to the lowest quantum state of the system 
$(n=0)$, introduces a length scale whose origin will be made clear soon. Our 
objective here is to derive an exact semiclassical trace formula for the density 
of states corresponding to the spectrum given by Eq.\ (\ref{spectrum}). This will 
enable us to identify the action of a single periodic orbit that generates the 
above spectrum. For an energy spectrum governed by only one quantum number, there 
is a rather simple way of deriving a trace formula. Following \cite{book} (Chapter 
3.2.2), we write $E_n=f(n)$, with degeneracy $D(n)=1$. The function $f(n)$ is 
monotonic with a differentiable inverse, $f^{-1}(x)=F(x)$. The exact density of 
states, defined by
\beq
g(E)=\Sigma_n\delta(E-E_n)\,, 
\label{dos}
\eeq
can then be rewritten, using Poisson resummation, as 
\beq   
g(E)=|F'(E)|\Sigma_{n=0}^{\infty}~\delta(n-F(E))\,=\,|F'(E)|\left[1+2\,
\Sigma_{k=1}^{\infty}\,\cos(2\pi kF(E))\right]\!.
\label{trace}
\eeq
This result, which is exact, can be split into two parts: a Thomas-Fermi (TF)
term $\widetilde{g}(E)=|F'(E)|$ which gives its average behaviour, and an 
oscillating term which we denote by $\delta g(E)$. 

In the semiclassical POT \cite{gutz}, the oscillating part of the exact density 
of states of a quantum Hamiltonian is expressed as a sum over the periodic orbits
of the corresponding classical Hamiltonian:
\beq
\delta g(E)=\sum_{\Gamma} \sum_{k=1}^{\infty} {\cal A}_{\Gamma\!,k} 
\cos\!\left[\frac {k}{\hbar}\, S_{\Gamma}(E)-\sigma_{\Gamma\!,k}\,\frac{\pi}{2}\right]
\label{osc}
\eeq
The sum is over primitive periodic orbits ($k=1$), denoted by $\Gamma$, and 
their repetitions (harmonics) $k>1$. The amplitude ${\cal A}_{\Gamma\!,k}$ 
of a periodic orbit (assumed here to be isolated in phase space) 
depends on its primitive period and on its stability matrix. 
The phase factor $\sigma_{\Gamma\!,k}$ is called the Maslov index of the periodic
orbit. Comparing the simple trace formula (\ref{trace}) with the general form 
(\ref{osc}), we see that it has only one primitive periodic orbit $\Gamma$ with 
action $S(E)=2\pi\hbar F(E)$, and an amplitude $2F'(E)$ which is proportional 
to its period. The Maslov index is zero. Since the spectrum is for $s$ states
only, the action is that of the {\it radial motion} in a central potential.
For the geometric spectrum, inverting \eq{spectrum}, we have 
\beq
n(E)=\frac{s_0}{2\pi}\ln(E_0/E)=F(E)\,.
\label{nofE}
\eeq
Therefore the action $S(E)$ of the primitive orbit is given by 
\beq
S(E)=\hbar s_0 \ln (E_0/E)\,.
\label{active}
\eeq
Substituting $F(E)$ in Eq.\ (\ref{trace}), we obtain 
\beq
g(E)=\frac{s_0}{2\pi |E|} \left\{1+2\,\sum_{k=1}^{\infty} \cos[ks_0
     \ln(E_0/E)]\right\}\!.
\label{trace2}
\eeq
This is an exact trace formula, representing a Fourier decomposition of \eq{dos}. 
When summed over all harmonics, it reproduces the quantized spectrum (\ref{spectrum}).

The smooth part of the density of states $\widetilde{g}(E)$ is given by the 
first term on the r.h.s.\ of (\ref{trace2}). The total number of Efimov states 
between the energies $-|E_0|$ and $-|E|$ is given by its integration over this 
interval, which yields Eq.\ \eq{nofE}.

Efimov trimers are formed when the two-body scattering length $|a|\gg r_0$, where 
the latter is of the order of the range of the intermolecular potential. 
One may then take the shallowest state to be $E_a\simeq \hbar^2\!/ma^2$, and 
the deepest state to have energy $E_0\simeq \hbar^2\!/mr_0^2$. As $a\rightarrow 
\pm \infty$, there is an accumulation of states near zero energy. Substituting 
these in \eq{nofE}, we find the total number of Efimov states to be 
\beq
N=\frac{s_0}{\pi} \ln (|a|/r_0)\,.
\label{numb2}
\eeq 
The same result was obtained by Efimov following a different route. 

In order to obtain a hint to the kind of one-body potential that can generate a spectrum 
of the form \eq{spectrum}, we apply an idea \cite{wusp} that exploits the properties
of the Abel transform \cite{abel}. The TF level density obtained above, interpreted
as the $s$-state level density of a radially-symmetric potential $V(r)$, can be
written \cite{book,rajat} as
\beq
\widetilde{g}(E) = -\frac{s_0}{2\pi E} 
                 = \sqrt{\frac{m}{2}}\frac{1}{\hbar\pi}\int_{r_1}^{r_2}\frac{dr}{\sqrt{E-V(r)}}\,,
\label{gtf}
\eeq
where $r_1$ and $r_2$ are the lower and upper turning points with $V(r_1) = E_0$
and $V(r_2)=E$, respectively, and the potential is assumed to be monotonously
increasing from its minimum value $E_0$ to the energy $E$. Using the substitution
$dr=y(V)\,dV$, so that $y(V)=1/V'(r)$, we can rewrite \eq{gtf} as
\beq
\widetilde{g}(E) =  \sqrt{\frac{m}{2}}\frac{1}{\hbar\pi}\int_{E_0}^E \frac{y(V)\,dV}{\sqrt{E-V}}\,.
\label{gev}
\eeq
The r.h.s.\ above represents an Abel transform \cite{abel} of the function $y(V)$ which, by
the inverse transform, is given by
\beq
y(V)=\hbar\sqrt{\frac{2}{m}}\left[\frac{\widetilde{g}(E_0)}{\sqrt{V-E_0}}
     +\int_{E_0}^{V}\widetilde{g}'(E)\,\frac{dE}{\sqrt{V-E}}\right].
\label{yV}
\eeq
Using $\widetilde{g}{\,'}(E)=s_0/2\pi E^2$, the integral above is elementary and leads to
\beq
y(V) = \frac{\hbar s_0}{\pi\sqrt{2m}}
       \left[\frac{1}{(-V)\sqrt{V-E_0}}+\frac{1}{(-V)^{3/2}}\arctan\sqrt{\frac{V-E_0}{(-V)}}\right].
\label{yVef}
\eeq
Assuming that $V(r)\to 0$ for $E\to 0$, the second term will be leading. We thus get asymptotically
\beq
y(V) \; \to \; \frac{\hbar s_0}{2\sqrt{2m}}(-V_{as})^{-3/2} = \frac{1}{V_{as}'(r)} \qquad 
               \qquad \hbox{for} \quad E\to 0\,.
\label{yVefas}
\eeq
(Note that both $E$ and $V$ are always negative.)
The potential $V_{as}(r)$ which solves the r.h.s.\ above is given by
\beq
V_{as}(r) = -\frac{\hbar^2s_0^2}{2mr^2}\,.
\label{Vasr}
\eeq
An inverse-squared potential of the type \eq{Vasr} is, indeed, shown in the following section
to be responsible for the asymptotic quantum-mechanical $s$-state spectrum of the Efimov 
three-body system in the limit $E\to 0$.

\section{The inverse-squared potential}
\label{secinv}

The three-body problem, after eliminating the center-of-mass coordinates, 
contains six degrees of freedom. In the hyperspherical formalism, these are 
described by a hyperradius 
\beq
R=\sqrt{(r_{12}^2+r_{23}^2+r_{31}^2)/3}
\eeq 
and five hyperangular coordinates \cite{bra}. In the adiabatic approximation, 
for fixed $R$, the Schr\"{o}dinger equation for the angular coordinates is 
solved to obtain a complete set of adiabatic eigenstates and the corresponding 
eigenvalues $\epsilon(R)$. In the unitary limit $a\rightarrow \pm\infty$, 
the angular variables decouple and one gets an effective inverse-squared 
potential in the coordinate $R$. The uncoupled hyperradial wavefunction in
the $L=0$ state is $\Psi(R)=R^{-5/2}u(R)$. The reduced wavefunction
$u(R)$ obeys the Schr\"odinger equation (cf.\ also a pedagogical review of the
Efimov effect \cite{rajat})
\beq
\left[-\frac{d^2}{d R^2}-\frac{(s_0^2+1/4)}{R^2} \right]\!
u(R)=\frac{2mE}{\hbar^2}\,u(R)\,,
\label{scheq}
\eeq
where $m$ is the mass of the atom and $s_0$ the constant given in \eq{space}.
In order to regularize the inverse-squared potential
\beq
V_0(R)=-\frac{\hbar^2}{2m}\frac{(s_0^2+1/4)}{R^2}\,,
\label{inverse}
\eeq     
we introduce a lower cut-off radius $R_c\simeq r_0$ as shown schematically in 
Fig.\ 1. 

\medskip

\begin{figure}[h]
\centering
\includegraphics[width=10cm]{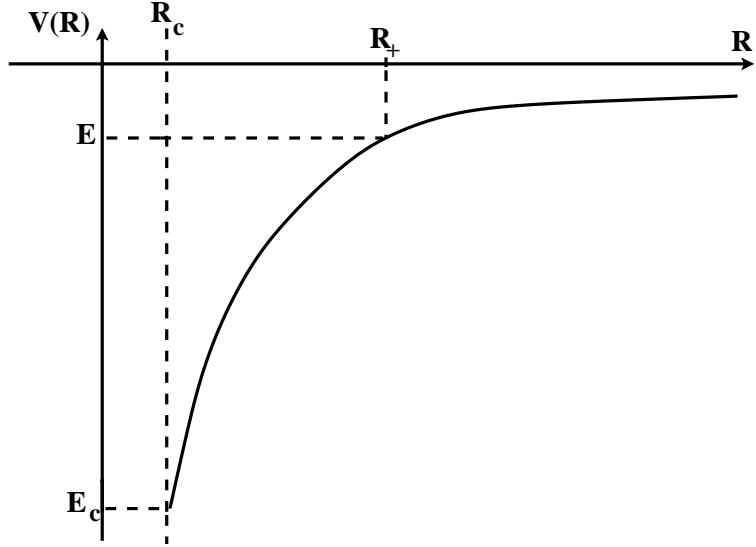}
\label{fig1}
\caption{A schematic of the inverse-squared potential is shown. $R_c$ is the 
cut-off radius; $R_+$ denotes the outer classical turning point which is 
determined by the energy through $E=V(R_+)$.}
\end{figure}

For future convenience we introduce a dimensionless scaled variable $x$ by
\beq
x=R/R_+\,, \qquad x_c = R_c/R_+\,,
\label{scales}
\eeq
where $R_+$ is the classical turning point at the energy $E$
\beq
E=V(R_+)=-\hbar^2 s_0^2/2mR_+^2\,,
\label{R+}
\eeq
(see Sec.\ \ref{secwkb} below), and the lowest energy $E_c$ of the classical 
particle is given by
\beq
E_c=V(R_c)=-\hbar^2 s_0^2/2mR_c^2\,,
\eeq
so that
\beq
x_c^2 = E/E_c\,.
\label{xcE}
\eeq

\subsection{Quantum-mechanical solutions}
\label{secqm}

We now solve the Schr\"{o}dinger equation \eq{scheq} with the lower boundary
condition $u(R_c)=0$. The second boundary condition comes from the requirement
that the wave functions vanish at infinity. We introduce the following transformation
\beq
u(R)= \sqrt{R}\,w(R)
\eeq
and obtain the equation for the function $w(R)$ as
\beq
\frac{d^2w}{d R^2}+\frac{1}{R}\frac{dw}{dR}+\frac{s_0^2}{R^2}\,w 
                  =-\frac{2mE}{\hbar^2}\, w\,.
\eeq
Using the scaled variable $x=R/R_+ = R\sqrt{-2mE/\hbar^2 s_0^2}$\,, 
the above equation reduces to the standard form for the modified Bessel 
function
\beq
\frac{d^2w}{d x^2}+\frac{1}{x}\frac{dw}{dx}-\frac{(is_0)^2}{x^2}\,w =w\,.
\label{Bes}
\eeq
Thus, the energy scales away, which is a unique feature of the inverse-squared 
potential whose $R$ dependence is the same as that of the kinetic energy operator.
The solution to the equation \eq{Bes}, given in \cite{rajeev}, is a modified Bessel 
function with imaginary index $is_0$
\beq 
w(x)=K_{is_0}(y)\,, \qquad y = s_0 x\,,
\eeq
which for real $y$ and $s_0$ is a real function that vanishes exponentially as $y\to\infty$.
The eigenstates are found from its zeros as explained below.

To compute the function $K_{i\nu}(y)$, we use a power series expansion given in \cite{da}:
\beq
K_{i\nu}(y)= -\sqrt{\frac{\nu\pi}{\sinh(\nu\pi)}}
              \sum_{k=0}^\infty \frac{(y^2\!/4)^k\sin{[\nu\ln(y/2)-\phi_{\nu,k}]}}
                                     {k!\,\nu\sqrt{\nu^2+1}\cdots\sqrt{\nu^2+k^2}}\,,
\label{Kx}
\eeq
where the phase $\phi_{\nu,k}$ is given by
\beq
\phi_{\nu,k}=\arg\Gamma(1+k+i\nu)=\phi_{\nu,0}+\sum_{s=1}^k \arctan\left(\frac{\nu}{s}\right).
\eeq
$\phi_{\nu,0}$ and a convergent series for its calculation \cite{abro} are given by 
\beq
\phi_{\nu,0}=\arg\Gamma(1+i\nu)
            =-\nu\,\gamma + \sum_{s=0}^\infty \left(\frac{\nu}{1+s}-\arctan\frac{\nu}{1+s}\right),
\eeq
where $\gamma=0.577215664\dots$ is Euler's constant. Numerically we find
\beq
\phi_{s_0,0}=-0.30103393\dots
\label{phi0}
\eeq

As stated in \cite{da}, the function $K_{i\nu}(y)$ has an infinite sequence of non-degenerate 
zeros $y_n$ ($n=1,2,\dots$) with $0 < \dots < y_{n+1}<y_n<y_{n-1}\dots<y_1<\nu$, and no zeros for 
$y\geq \nu$. The solution to the Schr\"{o}dinger equation (\ref{scheq}) is therefore given by
\beq
u(x) = C\sqrt{y}\,K_{i s_0}(y)\,, \qquad (y=s_0 x)
\label{ux}
\eeq
where $C$ is a normalisation factor. The eigenspectrum is obtained from the 
zeros $y_n$ via the boundary condition $K_{i\nu}(y_n)=0$ ($n=1,2,\dots$).

\begin{figure}
\vspace*{-0.6cm}
\includegraphics[width=16.5cm]{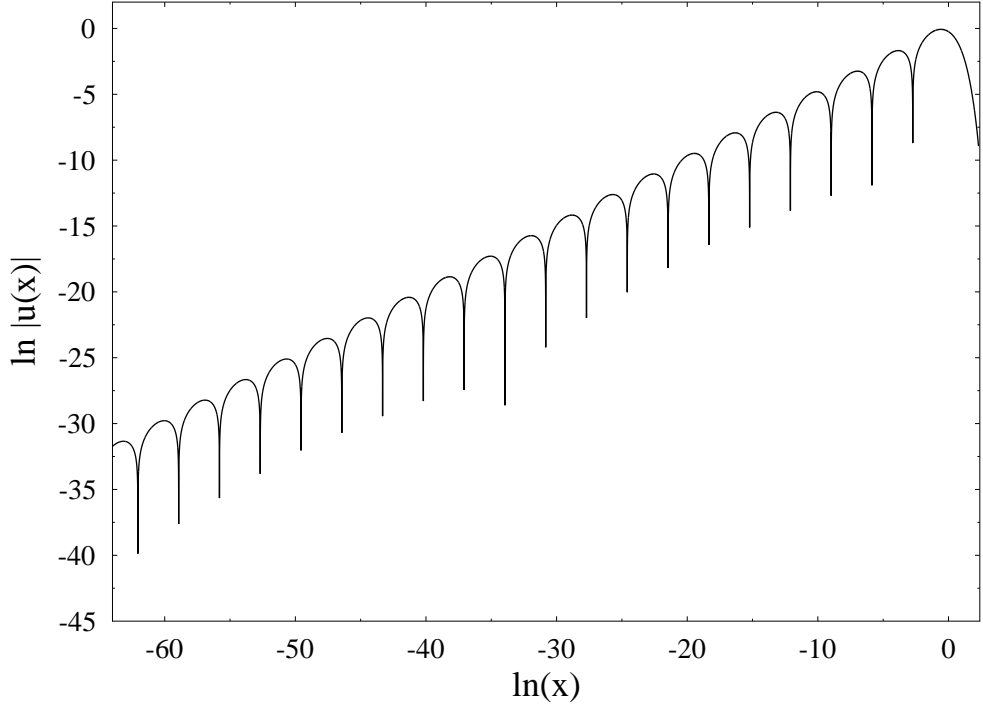}\vspace*{-1.8cm}
\label{fig2}
\caption{The universal function $u(x)$ in a doubly logarithmic plot. We see the first 20
zeros as negative spikes for $\ln(x)<0$.
}
\end{figure} 

It is an important feature of this system \cite{grif} that all eigenfunctions 
$u_n(x)$ are given in terms of the universal solution \eq{ux} simply by letting the 
variable $y$ in \eq{ux} start at ~the ($n$+1)-th zero $y_{n+1}$. Fig.\ 2 shows this 
universal function in a doubly logarithmic plot, exhibiting the first 20 zeros. 
For $x>1$, i.e.\ $\ln(x)>0$, we notice the beginning of the exponential tail.

We can now associate the zeros with eigenvalues of the scaled cut-off $x_c$
and write the eigenfunctions as
\beq
u_n(x) = C_n\sqrt{s_0 x}\,K_{i s_0}(s_0 x)\,, \qquad 
         x \geq (x_c)_n=y_{n+1}/s_0\,, \qquad n=0,1,\dots
\label{unx}
\eeq
where $C_n$ is the normalization constant of the $n$-th state. Note that $n$ here is the 
number of zeros $x_{n,j}>(x_c)_n$ ($j=0,1,\dots,n$) of the functions $u_n(x)$. We found 
numerically that the $C_n$ are practically identical for all $n\geq 2$ (see Tab.\ \ref{tab1} 
below), which is due to the fact that all wave functions are peaked around $x=1$ and the regions 
below the first three zeros, $y<y_3$, give only exponentially small contributions to the norms.

In the region of the maximum of $u(x)$ and the exponential tail for $y\simg1$, we had to 
include contributions to the sum in \eq{Kx} up to $k_{max}\simg 25$. For $y\simg 11$, we 
did not, in fact, obtain convergence of the $k$ sum. However, we found numerically that the 
leading term of an asymptotic form of $K_{i s_0}(y)$ for large $y$, given in \cite{da}
\beq
u_n(x) \;\sim\; C_n\sqrt{\pi/2}\,\exp(-y)\,,   \qquad   (y \gg 1)
\label{qtail}
\eeq
becomes sufficiently accurate for $y \simg 9$.

For $y \siml 0.3$, the terms with $k>0$ of the series in \eq{Kx} become numerically insignificant.
Since the largest zero is found at $y_1=0.0653423$ [cf.\ $\ln(x_c)_0=-2.73434$ in Tab.\ \ref{tab1} 
below], this means that all zeros of $K_{i\nu}(y)$ are given by the leading term with $k=0$, 
yielding the asymptotic solution given in \cite{bra}
\beq
K_{i\nu}(y) \; \propto \; \sqrt{y}\,\sin[\,\nu\ln(y/2)-\phi_{\nu,0}\,]\,. \qquad (n\gg 1)
\label{Kasy}
\eeq
Its zeros give the geometrical spectrum
\beq
E_n/\!E_c = (x_c)_n^2 \;\sim\; \left(\frac{2}{s_0}\right)^{\!\!2}\!
                               \exp\left(-2\pi\,n/s_0-2\pi/s_0+2\phi_{s_0,0}/s_0\right)  \qquad (n\gg 1)
\label{eratas}
\eeq
which, apart from constants, is the same as that discussed in Sec.\ \ref{secgeo};
in particular, the constant ratio $E_{n+1}/E_n$ remains the same. From \eq{eratas} we get 
\beq
\ln(x_c)_n \;\sim\; -n\pi/s_0 + \alpha_0\,, \qquad (n\gg 1)
\label{specas}
\eeq 
where $\alpha_0$ is, using the actual constants in \eq{space} and \eq{phi0}, given by
\beq
\alpha_0 = -\pi/s_0 + \phi_{s_0,0}/s_0 + \ln(2/s_0) = -2.73434\,. 
\label{alpha}
\eeq 

In column 1 of Tab.\ \ref{tab1}, we give some selected eigenvalues $\ln(x_c)_n$ obtained 
numerically from the exact solutions above. Although the expression \eq{specas} is mathematically
correct only asymptotically for large $n$, we find that the exact numerical eigenvalues agree
with the r.h.s.\ of \eq{specas} and \eq{alpha} within five digits -- which corresponds to our 
numerical accuracy -- even for $n=0$. 
\Table{tab1}{16}{
\begin{tabular}{|r|r|r|c|c|}\hline
$n$  & $\ln(x_c)_n$ {\footnotesize (QM)} &  $\ln(x_c)_n$ {\footnotesize (WKB)} & $\delta\ln(x_c)_n$ & $1/C_n$ \\ \hline
0   &   -2.73434   &   -2.64717  & 0.08717 & 0.3649953 \\
1   &   -5.85644   &   -5.77053  & 0.08591 & 0.3651889 \\
2   &   -8.97854   &   -8.89263  & 0.08591 & 0.3651892 \\
10  &  -33.95535   &  -33.86943  & 0.08592 & 0.3651892 \\
11  &  -37.07745   &  -36.99154  & 0.08591 & 0.3651892 \\
20  &  -65.17635   &  -65.09044  & 0.08591 & 0.3651892 \\
63  & -199.42668   & -199.34078  & 0.08590 & 0.3651892 \\
100 & -314.94440   & -314.85850  & 0.08590 & 0.3651892 \\ \hline
\end{tabular}}{ ~Eigenvalues $\ln(x_c)_n$ of some selected states, obtained numerically
both quantum-mechanically (QM) and in the WKB approximation discussed in Sec.\ \ref{secwkb}. 
Column 3 gives their differences, and column 4 shows the inverse normalization constants $C_n$ in \eq{unx}.}

\vspace{-0.3cm}
It is quite remarkable that, with the value $s_0\simeq 1$ given for the present system, the 
geometrical nature of the Efimov spectrum is preserved with a high numerical accuracy all 
the way down to $n=0$ in the inverse-squared potential. For appreciably larger values of 
$s_0$ this would not be the case.

We emphasize once more that the geometric nature of the asymptotic Efimov spectrum, given by
equations \eq{spectrum} and \eq{eratas}, is a direct consequence of the inverse-square 
behavior of the potential \eq{inverse}, with its strength determined by the parameter $s_0$. 
The ground-state energy $E_0$ in \eq{spectrum}, and with it the parameter $\phi_{s_0,0}$ in 
\eq{eratas}, is determined by the particular choice of the regularization of the potential
at small distances. In the present case, this was done by the hard-wall cut-off at $R_c$ and
the corresponding boundary condition. Other forms of regularization would lead to the same
geometric scaling as in \eq{spectrum} but to different ground-state energies $E_0$, such as
e.g.\ in \cite{mofri} where the potential \eq{inverse} was regularized by the addition of
a repulsive $1/R^4$ potential at short distances.


\subsection{WKB solutions}
\label{secwkb}

We now present our results using the WKB approximation. Since the WKB method is well
known, we need not present it here and refer for details to the text book by Migdal 
\cite{migdal}, who also discusses explicitly the Langer correction for radially symmetric
potentials \cite{lang}.

In the scaled variable $x$, the classical orbits are bounded inside the classically
allowed region $x_c \leq x \leq 1$. Applying the Langer correction with the potential 
\eq{inverse}, the effective classical potential becomes
\beq
V(R)=-\frac{\hbar^2}{2m}\frac{s_0^2}{R^2}\,, \qquad (R\geq R_c)
\label{VR}
\eeq
which is identical with the asymptotic potential \eq{Vasr} found from the inverse Abel transform
of the TF level density in Sect.\ II. The classical momentum then is
\beq
P(R)=\sqrt{2mE+\hbar^2s_0^2/\!R^2}\,,
\label{pr}
\eeq   
which, in the scaled variable $x$, becomes
\beq
P(R)=\frac{p(x)}{R_+}=\frac{p_0}{x}\sqrt{(1-x^2)}\,, 
     \quad p_0=\frac{\hbar s_0}{R_+}\,. \qquad  (x_c\leq x<1)
\label{Px}
\eeq   

The standard WKB wave function in the classically allowed region has the form
\beq
u_{in}^{\rm WKB}(x) = \frac{A}{\sqrt{p(x)}}\,
                      \cos\!\left[\frac{1}{\hbar}S_{in}(x)-\frac{\pi}{4}\right], \qquad (x_c\leq x < 1)
\label{WKBin}
\eeq
where $S_{in}(x)$ is the action integral along the orbit from $x$ to 1:
\beq
S_{in}(x)=\int_{x}^1\! p(x)dx=\hbar s_0\!\int_{x}^1 \frac{\sqrt{1-x^2}}{x}\,dx\,. \qquad (x_c\leq x < 1)
\eeq
This integral can be found analytically \cite{gr} and becomes
\beq
S_{in}(x) = \hbar s_0\! \left[\,\ln\left(\frac{1}{x}\right)
            +\ln\left(1+\sqrt{1-x^2}\right)
            -\sqrt{1-x^2}\,\right]\!. \qquad (x_c\leq x < 1)
\label{sin}
\eeq
The boundary condition that the wave function \eq{WKBin} vanishes at the lower
turning point $x=x_c$ leads, like in the quantum-mechanical case, to the quantization 
of the eigenenergies as shown below.

Outside the classically allowed region ($x>1$), the exponentially decaying WKB 
wave function has the form
\beq
u_{out}^{\rm WKB}(x)=\frac{B}{\sqrt{\kappa(x)}}\,\exp\left[-\frac{1}{\hbar}S_{out}(x)\right]\!,
                     \qquad \kappa(x)=\frac{\hbar s_0}{x}\,\sqrt{x^2-1}\,, \quad (x>1)
\label{WKBout}
\eeq
with the action $S_{out}(x)$ given by
\beq
S_{out}(x) = \!\int_1^x \!\!\kappa(x)\,dx
           = \hbar s_0\! \left[\arctan\left(\!\frac{1}{\sqrt{x^2-1}}\!\right)\!+\sqrt{x^2-1}
             -\frac{\pi}{2}\right]\!. \qquad (x>1)
\label{sout}
\eeq
For large $x\gg 1$, the function \eq{WKBout} has -- apart from the normalization -- the 
same exponential tail as the exact function \eq{qtail}:
\beq
u_{out}^{\rm WKB}(x) \;\sim\; \frac{B\,e^{\pi/2}}{\sqrt{s_0 \hbar}}\,\exp(-s_0 x)\,. \qquad  (s_0 x \gg 1)
\label{wkbtail}
\eeq
Equating the two asymptotic forms \eq{qtail} and \eq{wkbtail}, we can determine the
constant $B$ as
\beq
B_n = C_n\sqrt{s_0\hbar\,\pi/2}\,e^{-\pi/2},
\label{tailmatch}
\eeq
so that both wave functions agree exactly in the asymptotic tail region for each state $|n\rangle$.

Since both WKB functions \eq{WKBin} and \eq{WKBout} diverge at the upper classical turning 
point $x=1$, where $p(1)=\kappa(1)=0$, one must regularize these functions. This is done in
the standard ``connection'' \cite{lang,migdal} by a linear approximation to the potential $V(R)$ 
in the neighborhood of $R=R_+$ and matching the corresponding Airy function solution to the 
above WKB wave functions at some distances on either side of $x=1$. The requirement 
that the wavefunction be continuous and continuously differentiable at both matching points 
leads to the relation $A_n=2B_n$ for the normalization constants and the phase $-\pi/4$ in
\eq{WKBin}. 

We now have to fulfil the lower boundary condition at $x=x_c$:
\beq
u_{in}^{\rm WKB}(x_c) = 0 \quad \Rightarrow \quad
                          \cos\!\left[\frac{1}{2\hbar}\,S_0(E)-\frac{\pi}{4}\right] = 0\,,
\label{zeros}
\eeq
where $S_0(E)$ is the action of the primitive periodic orbit going from $x_c$ to $x=1$ and back:
\beq
S_0(E)=2\!\!\int_{x_c}^1 \!\!p(x)dx\,,
\eeq
which, using \eq{sin} and \eq{xcE} becomes
\beq
S_0(E)=\hbar s_0\!\left[\ln(E_c/E)
       +2\ln\left(1+\sqrt{1-E/E_c}\right)-2\sqrt{1-E/E_c}\right]\!.
\label{S0}
\eeq
For $|E|\ll|E_c|$, i.e.\ for the the shallow states ($n\gg 1$), the second and third terms above
become negligible and the leading term reproduces the action \eq{active} (with $E_0=E_c$) obtained 
from the periodic orbit theory applied to the geometric spectrum, as discussed in Sec.\ \ref{secgeo}.
Eq.\ \eq{zeros} has the solutions 
\beq
\frac{1}{2\hbar}\,S_0(E)-\frac{\pi}{4} = \left(\!n+\frac12\right)\pi\,, \qquad n=0,1,2,\dots
\eeq
which yields the WKB quantization condition
\beq
S_0(E_n^{\rm WKB}) = \oint P(R)\,dR = 2\pi\hbar\,(n+3/4)\,, \qquad n=0,1,2,\dots
\label{WKBquant}
\eeq
Note that the constant 3/4 in \eq{WKBquant} differs from the usual value 1/2, which one obtains 
for smooth potentials, due to the hard-wall reflection at the lower turning point $x_c$.

Using only the asymptotically leading logarithmic term in \eq{S0} yields the geometric
spectrum 
\beq
E_n^{\rm WKB} \;\sim\; E_c\,\exp[-2\pi/s_0(n+4/3)]\,, \qquad (n\gg 1)
\eeq
which corresponds to
\beq
\ln(x_c)_n \;\sim\; -n\pi/s_0 -3\pi/4s_0 = -n\pi/s_0 - 2.34158\,. \qquad (n\gg 1)
\eeq
This is the same as the quantum-mechanical result \eq{specas}, apart from a different shift 
(denoted by $\alpha_0$ there). In column 2 of Tab.\ \ref{tab1}, we give the WKB eigenvalues 
obtained from the quantization condition \eq{WKBquant}, using the full action \eq{S0}, in 
terms of the scaled values $\ln(x_c)_n$. We see that they come very close to the exact 
quantum values. In fact, there remains a slight shift in $\ln(x_c)_n$ that becomes constant 
for $n\geq 1$ (cf.\ the third column in Tab.\ \ref{tab1}). A similar (but different) shift 
between the asymptotic exact and WKB spectra has been obtained for an attractive $1/R^2$ 
potential regularized differently \cite{mofri} (see \cite{fried} for its interpretation).

We now have to regularize the WKB functions \eq{WKBin}, \eq{WKBout} near the turning point
$(x=1)$ where they diverge. While the standard connection to the Airy solution of the locally 
linearized potential leads to the WKB quantization condition \eq{WKBquant}, as described
above, we found that it does not yield satisfactory wave functions. The reason is that the
asymptotic Airy solutions do not come sufficiently close to the WKB solutions on either side 
of the turning point. We were, however, successful when using Langer's generalized 
``connection formula'' for one isolated classical turning point [Ref.\ \cite{lang}, Eqs.\ 
(11a) and (11b) with $\eta=0$]. Expressable in terms of a single Airy function Ai$(\xi)$, it reads
\beq
u_n^{uni}(x) = D_n\,\sqrt{{\cal S}(x)/|\xi|}\,{\rm Ai}(\xi)\,,\qquad x \geq (x_c)_n
\label{uuni}
\eeq 
where $D_n$ is a normalization constant, and
\begin{eqnarray}
\xi & = & \left[\frac32\,\frac{1}{\hbar}\,S_{out}(x)\right]^{2/3},\quad
          ~{\cal S}(x) = \frac{S_{out}(x)}{\hbar \kappa(x)}\qquad\!\hbox{for}  
          \quad  x\geq 1\,,\;\;\xi \geq 0\,,\label{xiout}\\
\xi & = & -\left[\frac32\,\frac{1}{\hbar}\,S_{in}(x)\right]^{2/3},\quad
          {\cal S}(x) = \frac{S_{in}(x)}{\hbar p(x)}\qquad\hbox{for}   
          \quad  x\leq 1\,,\;\;\xi \leq 0\,.\label{xiin}
\end{eqnarray}
The superscript ``uni'' in \eq{uuni} stands for ``uniform'', because it turns out that
\eq{uuni} is a global uniform approximation that can be used not only in the vicinity
of the classical turning point, but throughout the whole domain $(x_c)_n \leq x < \infty$.
By construction \cite{lang}, it yields the asymptotic WKB solutions \eq{WKBin}, \eq{WKBout} 
sufficiently far from the turning point $x=1$.

\begin{figure}
\vspace*{-0.8cm}
\begin{center}
\includegraphics[width=15.3cm]{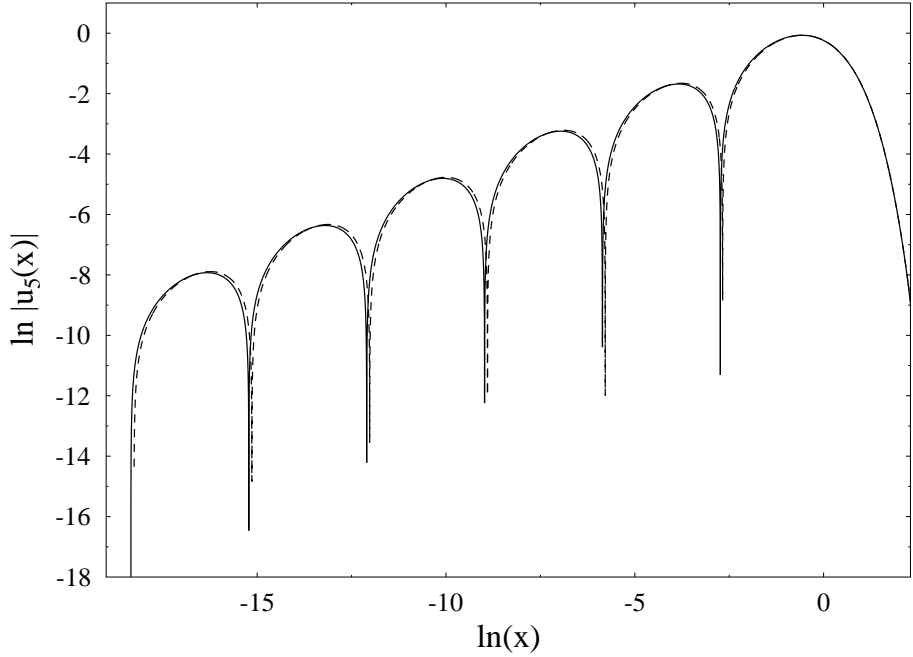}\vspace*{-1.4cm}
\end{center}
\label{fig3}
\caption{The state $u_5(x)$ in a doubly logarithmic plot. Solid line: exact normalized 
quantum-mechanical solution \eq{unx}. Dashed line: normalized uniform WKB approximation \eq{uuni}.
}
\end{figure} 

The normalized uniform WKB wavefunction \eq{uuni} for $n=5$ is shown in Fig.\ 3, along with the 
exact one, in a doubly logarithmic plot. Apart from the slightly shifted zeros, there is a 
surprisingly good agreement. 
Fig.\ 4 shows the ground-state wave function for $n=0$, both exactly (solid line), the
``raw'' WBK approximation (dotted line), and in the uniform approximation \eq{uuni} (dashed
line). Even for this lowest state, the quantum-mechanical and the uniform WKB solutions
are practically indistinguishable.

\bigskip

{\it Note added after publication of this paper:}

\medskip

The above global uniform approximation is also discusse in:\\
C. M. Bender and S. A. Orszag: {\it Advanced Mathematical Methods for Scientists and Engineers}
(Springer-Verlag New York 1991), Ch.\ 10.4, pp. 510 ff.

\begin{figure}
\vspace*{-0.8cm}
\begin{center}
\includegraphics[width=15.3cm]{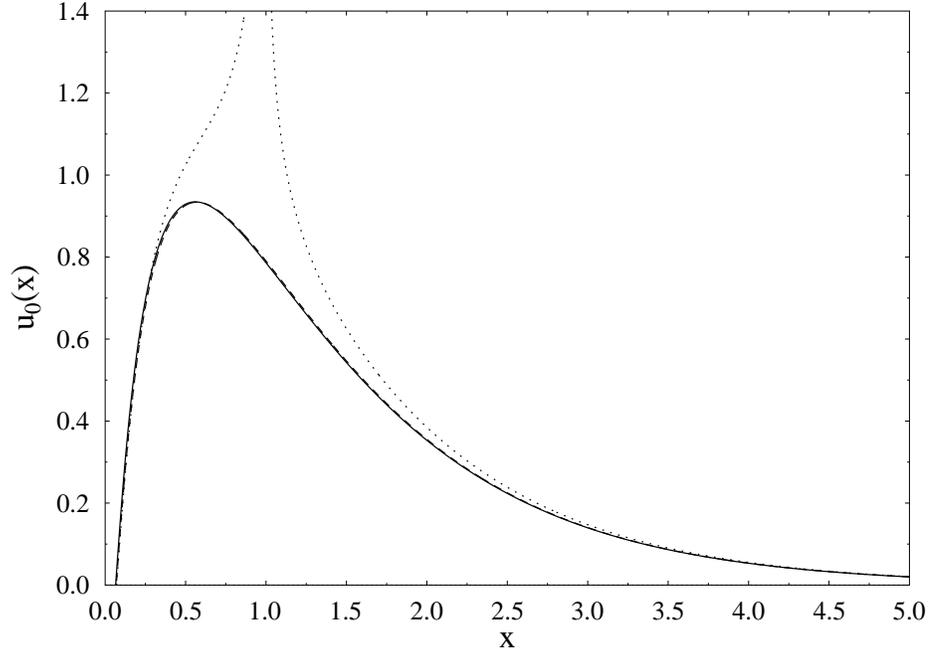}\vspace*{-1.4cm}
\end{center}
\label{fig4}
\caption{The ground state $u_0(x)$ in a linear plot. Solid line: exact normalized 
quantum-mechanical solution \eq{unx}. Dotted line: WKB approximation \eq{WKBin}, \eq{WKBout}, 
normalized to the exact solution in the tail region via equation \eq{tailmatch}. 
Dashed line: normalized uniform WKB approximation \eq{uuni}.
}
\end{figure} 

\newpage

\subsection{Mean squared radii}

The size of our system in the $n$-th state is given by the mean squared hyperradius defined as
\beq
\langle\, R^2\, \rangle = \frac{\langle\Psi_n|R^2|\Psi_n\rangle}{\langle\Psi_n|\Psi_n\rangle}
                    = R_+^2(E_n)\int_{x_c}^{1} dx\,x^2\,|u_n(x)|^2,
\eeq
where Eq.\ \eq{scales} and the normalized wavefunctions $u_n(x)$ have been used. Since $R_+^2$ 
scales like $1/E_n$, see \eq{R+}, the ratio of mean squared radii of two successive states is 
$E_n/E_{n+1}$, if the expectation values  $\langle n|x^2|n\rangle$ remain independent of $n$. 

That this is, indeed, the case for all $n\geq 1$, is demonstrated in Tab.\ \ref{tab2}, both for 
the quantum-mechanical and the uniform WKB solutions. Thus the mean squared radius increases by 
the same geometric scaling factor as the energy decreases, as discussed in \cite{bra}, also 
in the WKB approximation.

\Table{tab2}{13.5}{
\begin{tabular}{|r|c|c|}\hline
$n$\; & $\langle n|x^2|n\rangle$ {\footnotesize (QM)} & $\langle n|x^2|n\rangle$ {\footnotesize (WKB)} \\ \hline
\;0~  & 1.3265  & 1.3408 \\
\;1~  & 1.3251  & 1.3392 \\
\;2~  & 1.3251  & 1.3392 \\
\;5~  & 1.3251  & 1.3392 \\
\;10~ & 1.3251  & 1.3392 \\ \hline
\end{tabular}}{ Mean squared radii $\langle n|x^2|n\rangle$ of some  
states, obtained quantum-mechanically (QM) and in the uniform WKB 
approximation. 
}

\newpage

\section{Summary}

To summarize, we have shown that the ($s$-wave) geometric spectrum of the Efimov 
energy levels in the unitary limit is semiclassically generated by a single 
periodic orbit whose action depends logarithmically on the energy. 
The smooth part of the $s$-state level density, obtained by the periodic
orbit theory, is consistent with an attractive inverse-squared radial
one-body potential. We have re-derived the quantum spectrum of the 
inverse-squared potential with a lower cut-off and shown, for $s_0\simeq 1$, 
that it reproduces the geometric Efimov spectrum not only for shallow states, 
but yields the same constant ratio $E_{n+1}/E_n$ all the way down to the ground 
state with $n=0$. We have given an analytical expression for the zeros of 
the eigenfunctions. The WKB quantization of the classical system (including 
the Langer correction) yields the same spectrum, although slightly phase 
shifted, which preserves the same constant ratio $E_{n+1}/E_n$ down to $n=0$. 
The action of the classical system has as its leading term precisely the 
action obtained from the periodic orbit theory. The WKB wave functions,
when regularized around the classical turning point using Langer's generalized
connection formula, reproduce the exact ones surprisingly well, apart from the 
slightly shifted zeros -- even for the ground state ($n=0$). Both the quantum 
and the WKB solutions reproduce the geometric scaling of the mean squared radii 
of the Efimov states which is inverse to that of their eigenenergies.

As mentioned in the introduction, the Efimov spectrum is experimentally not 
measured in the unitary limit, but for finite and large scattering lengths 
$|a|\gg r_0 $. Strictly speaking, therefore, the $1/R^2$ nature of the effective 
potential is only guaranteed in the range $R_c \leq R \siml |a|$. Then, the 
geometric scaling of the spectrum would only hold within the corresponding energy 
range, the approximate number of Efimov bound states being given by Eq.\ \eq{numb2}. 
Also, the derivation of the inverse-squared potential given in Sec.\ II and the 
analytical exact wavefunctions given in Sec.\ III A would no longer hold. However, 
once we assume an inverse-squared potential in the range $R_c \leq R \leq |a|$, 
our WKB calculations still go through provided we restrict the outer turning point 
to $R_+ \leq |a| $.

\bigskip

R.K.B.\ and M.B.\ are grateful to the IMSc, Chennai, for its hospitality,
excellent working conditions and financial support. We are also grateful to an
anonymous referee for valuable suggestions. M.B.\ acknowledges 
enlightening discussions with H. Friedrich, C. Eltschka and D. L. Sprung.

\end{document}